# Pure spin current in a robust pigment-red film

Kazuhiro Nishida,[1] Yoshio Teki,[2] and Eiji Shikoh[1,a)]

[1]*Graduate School of Engineering, Osaka City University, 3-3-138 Sugimoto, Sumiyoshi-ku, Osaka 558-8585, Japan*

[2]*Graduate School of Science, Osaka City University, 3-3-138 Sugimoto, Sumiyoshi-ku, Osaka 558-8585, Japan*

We report the spin current properties in a pigment-red (perylene-3,4,9,10-tetracarboxylic dianhydride: PTCDA) film prepared by thermal evaporation. In a palladium(Pd)/PTCDA/$Ni_{80}Fe_{20}$ tri-layer sample, a pure spin-current is generated in the PTCDA layer by the spin-pumping of the $Ni_{80}Fe_{20}$. The spin current is absorbed into the Pd layer, converted into a charge current with the inverse spin-Hall effect in Pd, and detected as an electromotive force. This is clear evidence for the pure spin current in a PTCDA film, and it is confirmed that a PTCDA film is useful not only as a robust protection layer material but also as a spintronic material.

[a)]E-mail: shikoh@eng.osaka-cu.ac.jp

Carbon-based molecular materials have attracted much attention for the wide possibility of spintronic applications. [1-17] Because of the good spin coherence originating from the weak spin-orbit interaction in the materials, it is expected to switch the spin current in the molecular materials through the visible light irradiation, by utilizing the photoconductivity in molecular materials. [7-9] It is also expected to produce spin-related functions through the visible light irradiation, by controlling the excited $\pi$-electron states in molecular materials. [10] Moreover, to control the spin transport property by applying a pressure may be realized by utilizing the flexibility and the spin-orbit interaction differences due to the molecular structural difference. [11] In these attractive examples, improvement of the physical durability of the molecular materials is an indispensable issue on development of the molecular spintronics as similar to the case of the conventional molecular electronics.

To overcome the above physical durability problem in the conventional $\pi$-conjugated molecular electronics, a copper-phthalocyanine (CuPc) molecular film [13-16] and a perylene-3,4,9,10-tetracarboxylic dianhydride (PTCDA) molecular film [16-18] are generally used. The former molecule is called as a pigment-blue and the latter is called as a pigment-red, which are robust materials due to the rigid molecular frames. These $\pi$-conjugated molecular films are easily formed by thermal evaporation in a vacuum, and PTCDA films are known as durable even in high-dense plasma like the sputtering method which is a standard technique to form thin films. [18] Thus, even in spintronics, CuPc and PTCDA films will probably be useful by the similar reason on conventional molecular electronics. The spin physics and application related to CuPc films has already been well-investigated, [13-16] meanwhile those of PTCDA molecular films have not been studied much yet. [16,17] Previously, with a magnetic tunnel junction structure, the spin transport in PTCDA films using a spin-polarized current has been reported, [17] where the PTCDA film thickness was only 2 nm. This means that the spin transport mechanism in PTCDA films is still unclear because there are two possibilities of the spin transport mechanism [17]: one is due to a tunneling process through the PTCDA film (not a spin injection process), and another is due to a spin-injection process into the PTCDA films. Thus, to clarify the spin transport mechanism in PTCDA molecular films is necessary for future spintronic application. In this study, we show the spin current, which is a flow of spin angular momenta, generated in thermally-evaporated PTCDA films at room temperature (RT) by using a pure spin current induced by the spin-pumping. [4-9,11,19-23] This is clear evidence for a pure spin current in PTCDA films via a spin injection process, and it is confirmed that a PTCDA film is useful not only as a robust protection layer material but also as a spintronic material.

Our sample structure and experimental set up are illustrated in Figure 1. Spin transport in a PTCDA film is observed as follows: in palladium(Pd)/PTCDA/$Ni_{80}Fe_{20}$ tri-layer samples, a spin-pump-induced pure spin current, $\vec{J}_S$, driven by ferromagnetic resonance (FMR) [19,20] of the $Ni_{80}Fe_{20}$ film is generated in the PTCDA layer. This $\vec{J}_S$ is then absorbed into the Pd layer. The absorbed $\vec{J}_S$ is converted into a charge current as a result of the inverse spin-Hall effect (ISHE) [22] in Pd and detected as an electromotive force, $\vec{E}$, [4-9,11,21-23] which is expressed as,

$$\vec{E} \propto \theta_{SHE} \vec{J}_S \times \vec{\sigma}, \qquad (1)$$

where $\theta_{SHE}$ is the spin-Hall angle which is the conversion efficiency from a spin current into a charge current in the material, and $\vec{\sigma}$ is the spin-polarization vector in the $\vec{J}_S$. That is, if electromotive force due to the ISHE in Pd is detected under the FMR of the $Ni_{80}Fe_{20}$, it is clear evidence for spin transport in a PTCDA film.

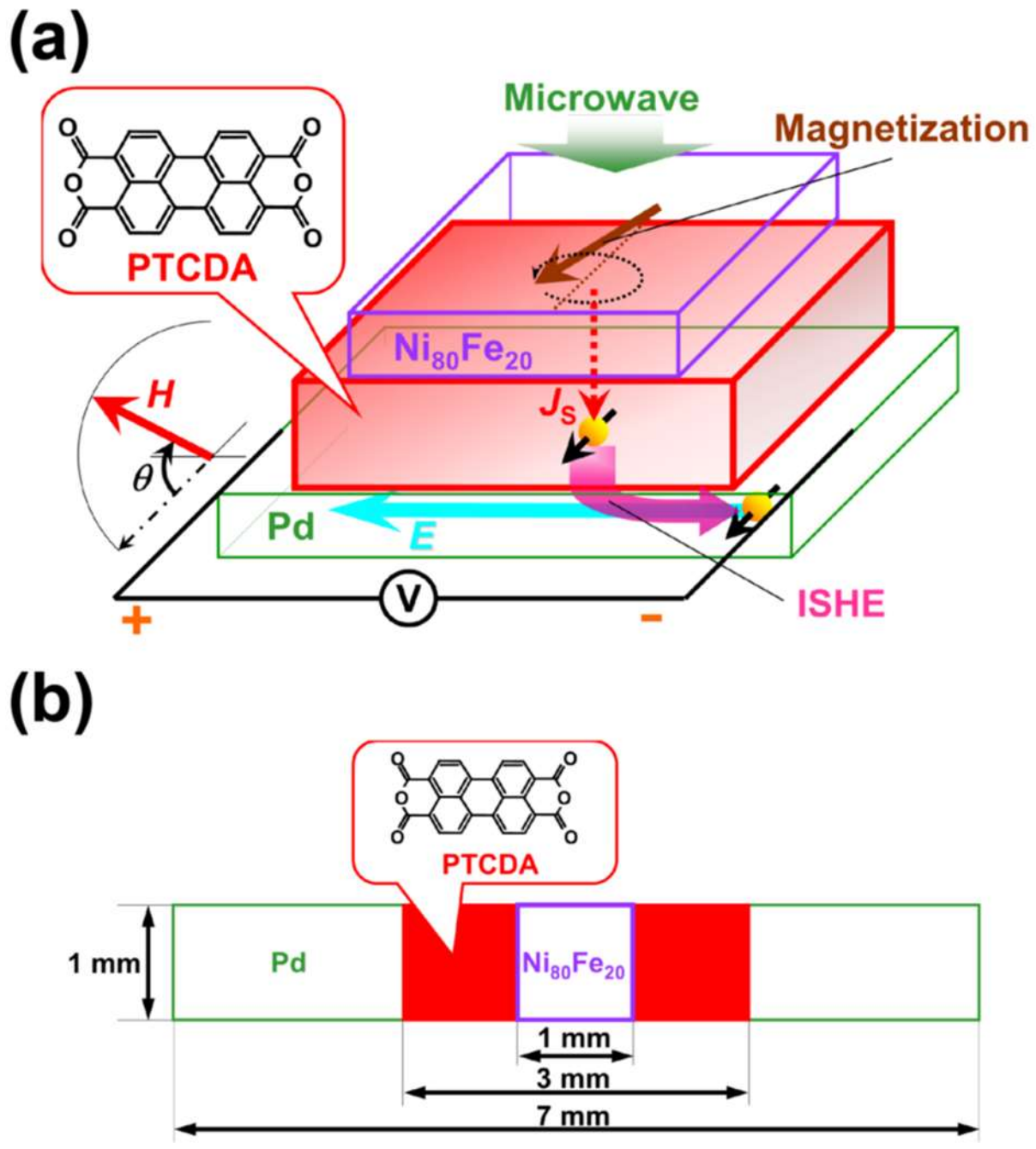

FIG. 1. (a) Bird's-eye-view and (b) top-view illustrations of our sample and an orientation of external applied magnetic field $\vec{H}$ with an angle ($\theta$) to the sample film plane used in the experiments. $\vec{J}_S$ and $\vec{E}$ correspond, respectively, to the spin current generated in the pentacene film by spin-pumping and the electromotive forces due to the ISHE in Pd.

Electron beam (EB) deposition technique was used to deposit Pd (Furuuchi Chemical Co., Ltd., 99.99% purity) to a thickness of 10 nm on a thermally-oxidized silicon substrate, under a vacuum pressure of $<10^{-6}$ Pa. Next, also under a vacuum pressure of $<10^{-6}$ Pa, PTCDA molecules (Tokyo Chemical Industry Co., Ltd.; sublimation grade, >98%) were thermally evaporated through a shadow mask. The deposition rate and the substrate temperature during PTCDA depositions were set to 0.1 nm/s and RT, respectively. The PTCDA layer thickness (*d*) was varied between 15 and 60 nm. Finally, $Ni_{80}Fe_{20}$ (Kojundo Chemical Lab. Co., Ltd., 99.99%) was deposited by EB deposition through another shadow mask, under a vacuum pressure of $<10^{-6}$ Pa. During $Ni_{80}Fe_{20}$ depositions, the sample substrate was cooled with a cooling medium of -2°C, to prevent the deposited molecular films from breaking. As a control experiment, samples with a Cu layer, instead of the Pd layer, were prepared.

We used a microwave $TE_{011}$-mode cavity in an electron spin resonance system (JEOL, JES-TE300) to excite FMR in $Ni_{80}Fe_{20}$, and a nano-voltmeter (Keithley Instruments, 2182A) to detect electromotive forces from the samples. Leading wires for detecting the output voltage properties were directly attached with silver paste at both ends of the Pd (or Cu) layer. All of the measurements were performed at RT.

Figure 2(a) shows the FMR spectrum of a sample with a Pd layer and with the *d* of 30 nm at an external magnetic field orientation angle ($\theta$) to the sample film plane of 0°, under a microwave power of 200 mW. The FMR field ($H_{FMR}$) of the $Ni_{80}Fe_{20}$ film is 1,049 Oe and the $4\pi M_s$ of the $Ni_{80}Fe_{20}$, where $M_s$ is the saturation magnetization of the $Ni_{80}Fe_{20}$ film, is estimated to be 8,806 G at a microwave frequency ($f$) of 9.45 GHz and under FMR conditions in the in-plane field:

$$\frac{\omega}{\gamma} = \sqrt{H_{FMR}(H_{FMR} + 4\pi M_S)}, \qquad (2)$$

where $\omega$ and $\gamma$ are the angular frequency ($2\pi f$) and the gyromagnetic ratio of $1.86\times10^7$ $G^{-1}s^{-1}$ of $Ni_{80}Fe_{20}$, respectively. [7,21,23] Fig. 2(b) shows the output voltage properties of the same sample shown in Fig. 2(a); the circles represent experimental data and the solid lines are the curve fit obtained using the equation [7,21-23]:

$$V(H) = V_{Sym}\frac{\Gamma^2}{(H-H_{FMR})^2+\Gamma^2} + V_{Asym}\frac{-2\Gamma(H-H_{FMR})}{(H-H_{FMR})^2+\Gamma^2}, \qquad (3)$$

where $\Gamma$ denotes the damping constant (56 Oe in this study). The first and second terms in eq. (3) correspond to the symmetry term to $H$ due to the ISHE, and the asymmetry term to $H$ due to the anomalous Hall effect and other effects showing the same asymmetric voltage behavior relative to the $H$ like parasitic capacitances, respectively. [7,21-23] $V_{Sym}$ and $V_{Asym}$ correspond to the coefficients of the first and second terms in eq. (3). Output voltages are observed at $H_{FMR}$ at $\theta$ of 0 and 180°. Notably, the output voltage changes sign between $\theta$ values of 0 and 180°. This sign inversion of voltage in Pd associated with the magnetization reversal in $Ni_{80}Fe_{20}$ is characteristic of ISHE. [7,21-23]

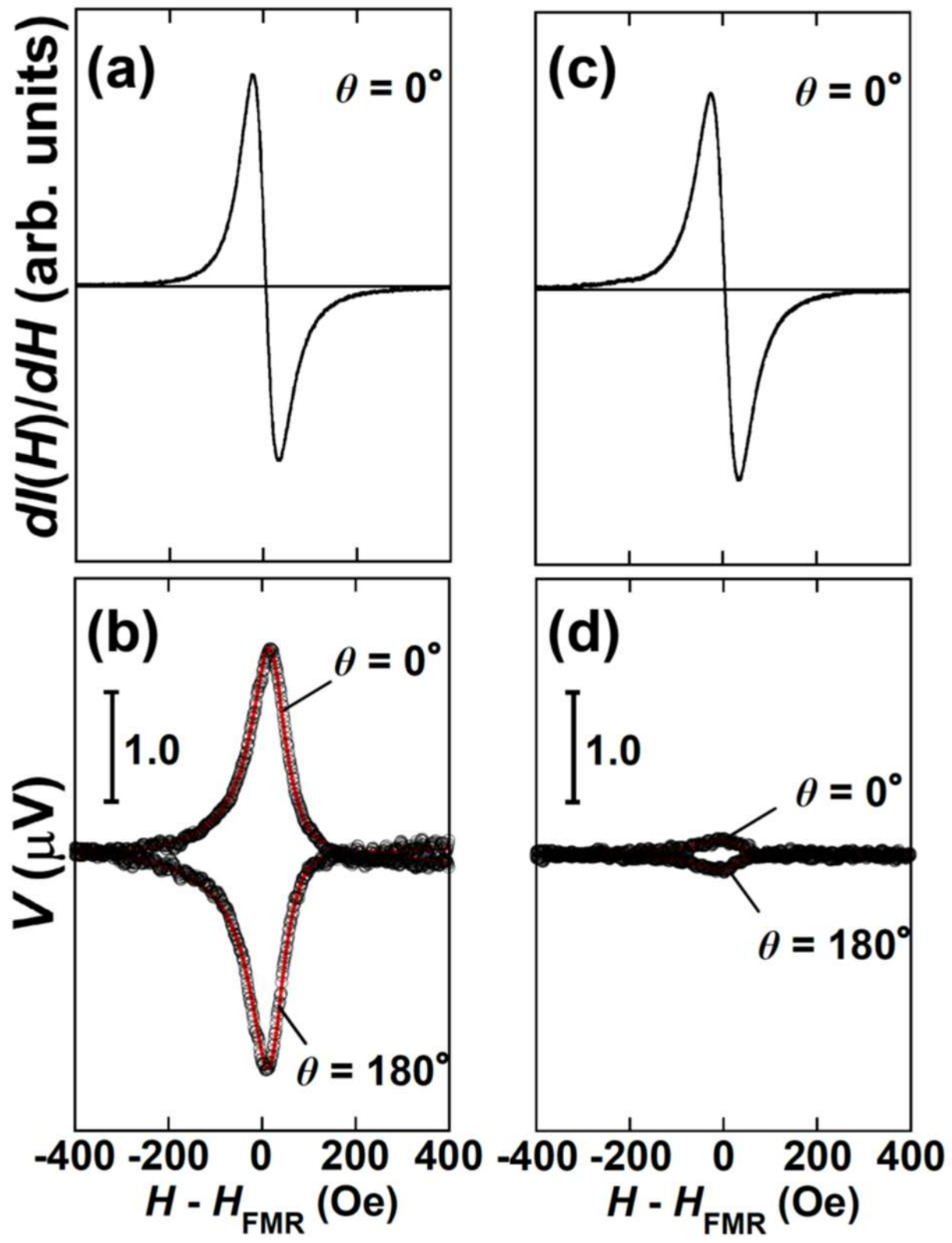

FIG. 2. (a) FMR spectrum and (b) output voltage properties of a sample with a Pd layer. (c) FMR spectrum and (d) output voltage properties of a sample with a Cu layer. In both cases, the PTCDA film thickness is 30 nm and the applied microwave power is 200 mW.

As a control experiment, we tested samples with a Cu layer, where the spin-orbit interaction is relatively weak, instead of the Pd layer. Fig. 2(c) shows the FMR spectrum of a sample with a Cu layer and with the $d$ of 30 nm at the $\theta$ of 0°, under a microwave power of 200 mW. Fig. 2(d) shows output voltage properties of the same sample as shown in Fig. 2(c), where very small electromotive force compared with Fig. 2(b) was observed at $\theta$ values of 0 and 180°. This comes from the $\theta_{SHE}$ difference between Pd and Cu due to the spin-orbit interaction difference. As another control experiment, we studied the microwave power ($P$) dependence of the electromotive forces in a sample with a Pd layer; the results are shown in Fig. 3. The $V_{Sym}$ increases in proportion to the increase in $P$. The above results suggest that the electromotive forces at the FMR field ($H - H_{FMR} = 0$) observed for the sample with a Pd layer (see Fig. 2(b)) are mainly due to the ISHE in Pd, that is, spin-pump-induced spin transport has been achieved in an evaporated PTCDA film at RT. This is clear evidence for a spin current in PTCDA films via a spin injection process.

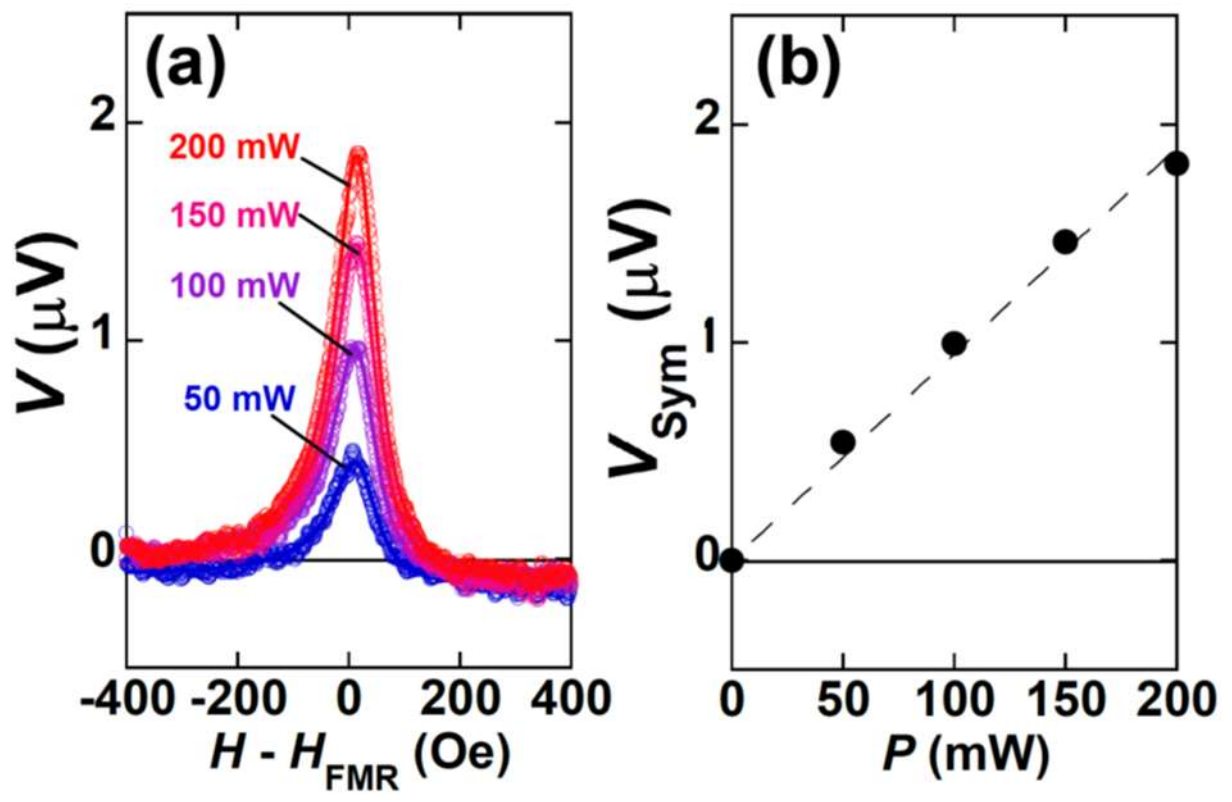

FIG. 3. (a) Microwave power ($P$) dependence of electromotive force and (b) an analysis result obtained with eq. (3). $V_{Sym}$ corresponds to the coefficient of the first term in eq. (3). The PTCDA film thickness is 30 nm. The dashed line in (b) is a linear fit.

Figure 4 shows the $d$ dependence of (a) $4\pi M_s$ calculated via eq. (2) and of (b) $V_{ISHE}$ estimated via eq. (3). With increasing $d$, $M_s$ decreases slightly, while $V_{ISHE}$ clearly decreases. We estimated the spin diffusion length of the PTCDA film ($\lambda_s$) to be ~30 nm at RT under the assumption [4-7,21] of an exponential decay of the spin current in the PTCDA film. The dotted line in Fig. 4(b) represents the result of this estimation.

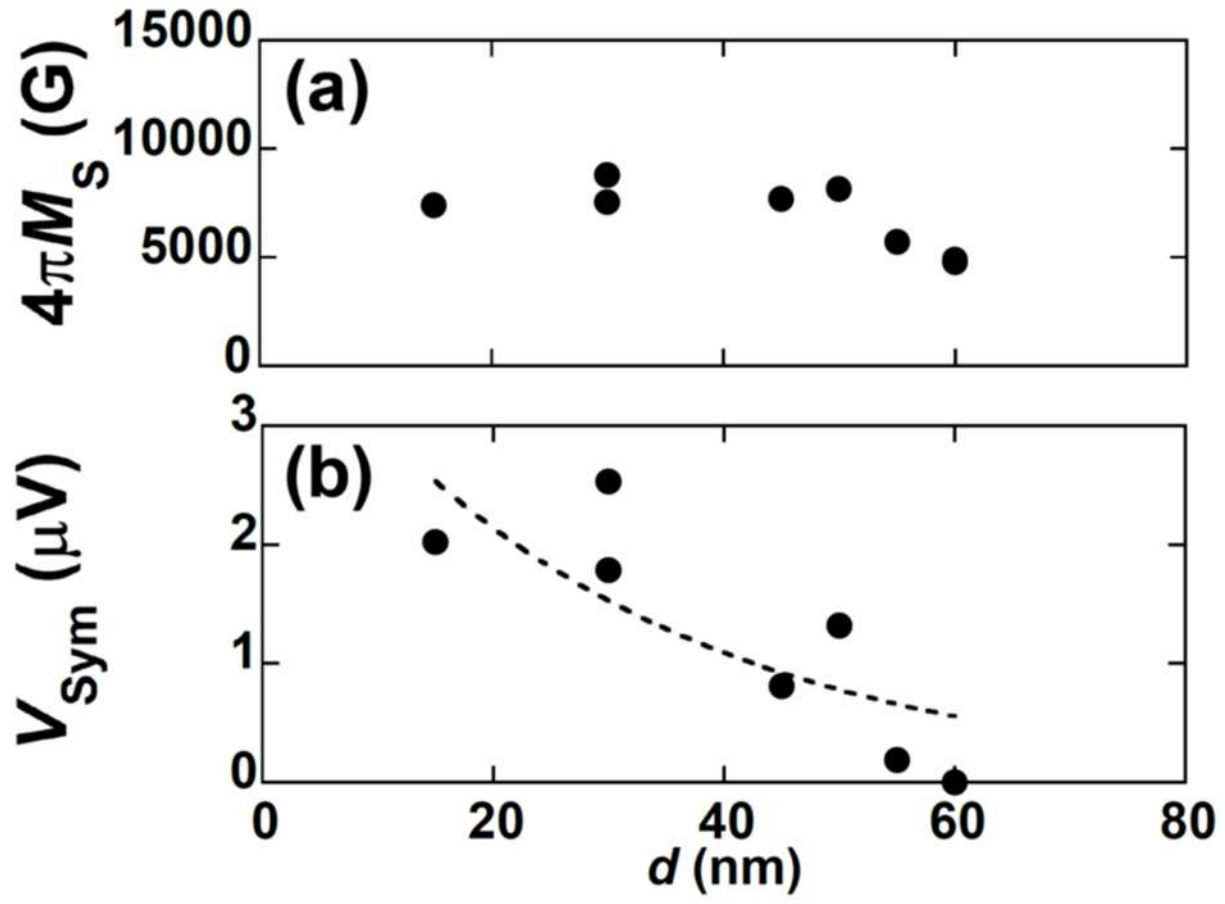

FIG. 4. Dependence of (a) $4\pi M_s$ ($M_s$: saturation magnetization), calculated by eq. (2), and of (b) $V_{Sym}$ estimated by eq. (3), on the PTCDA film thickness (*d*). Circles are the experimental data. The dotted line in (b) is a curve fit under the assumption of exponential decay.

We discuss the validity of the $\lambda_s$ estimation of our PTCDA films with the reported values in other molecular materials studied at RT by using a spin-pump-induced spin current. [4-9,11] The estimated $\lambda_s$ value, ~30 nm in PTCDA films at RT is comparable to the reported spin diffusion lengths of other thermally-evaporated molecule films: ~13 nm for $C_{60}$ fullerene, [11] ~25 nm for $C_{84}$ fullerene, [11] ~35 nm for TIPS-pentacene, [9] ~42 nm for pentacene, [7] and ~50 nm for $Alq_3$. [5] Polymer films, such as PBTTT and PEDOT:PSS films, tend to possess longer spin diffusion lengths than low-molecular-weight molecular films: ~150 nm for PBTTT, [4] and ~150 nm for PEDOT:PSS. [6] It is reported that the surface morphology on a PTCDA film is much rougher than an $Alq_3$ film. [16]

This must cause the penetration of Ni and Fe atoms into PTCDA films more than the cases of other molecular films during the sample preparation processes. Also, in a previous case of pentacene films, [7] the film crystal structure is a factor to decide the spin diffusion length, that is, the higher the crystallinity of a molecular film is, the longer the spin diffusion length in the molecular film might be. On the other hand, $\pi$-conjugated electron numbers per molecule are not affected to decide the spin diffusion length in molecular films. Thus, for low-molecular-weight molecular films, to improve the film crystallinity is an indispensable issue to obtain longer the spin diffusion length, although the spin diffusion length of several tenth nm is enough for spintronic application because the recent nanotechnology has been evolved much.

Finally, we compare our pigment-red PTCDA study with a pigment-blue CuPc study. [13] The spin transport distance in a CuPc film is estimated to be ~1 nm in pure spin injection process and ~35 nm in contaminated spin injection process through an inelastic process by using a two-photon photoemission process. [13] Although the evaluation process is different, the spin diffusion lengths in PTCDA and CuPc films have been estimated comparable. Thus, when the two kind pigments are used as physically durable and spin-transporting materials, the two pigments can separately be used on the viewpoint of an optical reason, and it is probably same as the case of conventional molecular electronics.

In summary, spin transport properties of thermally-evaporated pigment-red PTCDA films were studied at RT by using the spin-pumping for the spin injection and using the ISHE in non-magnetic metals for the spin detection methods. We achieved spin transport in PTCDA films; the spin diffusion length in PTCDA films was estimated to be about 30 nm at RT. This is clear evidence for the pure spin current in PTCDA films prepared by thermal evaporation, and it is confirmed that a PTCDA film is useful not only as a protection layer material but also as a spintronic material, which paves the way to developing molecule-based spintronic devices.

This research was partly supported by the JSPS Grants-in-Aid for Scientific Research (B) (No. 16H04136 (to Y. T.)), and by the Osaka City University (OCU) Strategic Research Grant 2018 for top priority researches (to E. S.).

References

[1]M. Shiraishi and T. Ikoma, Physica E **43**, 1295 (2011).

[2]Z.H. Xiong, Di Wu, Z.V. Vardeny, and J. Shi, Nature **427**, 821 (2004).

[3]M. Ohishi, M. Shiraishi, R. Nouchi, T. Nozaki, T. Shinjo, and Y. Suzuki, Jpn. J. Appl. Phys. **46**, L605 (2007).

[4]S. Watanabe, K. Ando, K. Kang, S. Mooser, Y. Vaynzof, H. Kurebayashi, E. Saitoh, and H. Sirringhaus, Nature Phys. **10**, 308 (2014).

[5]S.W. Jiang, S. Liu, P. Wang, Z. Z. Luan, X. D. Tao, H. F. Ding, and D. Wu, Phys. Rev. Lett. **115**, 086601 (2015).

[6]M. Kimata, D. Nozaki, Y. Niimi, H. Tajima, and Y. Otani, Phys. Rev. B **91**, 224422 (2015).

[7]Y. Tani, Y. Teki, and E. Shikoh, Appl. Phys. Lett. **107**, 242406 (2015).

[8]Y. Tani, T. Kondo, Y. Teki, and E. Shikoh, Appl. Phys. Lett. **110**, 032403 (2017).

[9]Y. Tanaka, T. Kono, Y. Teki, and E. Shikoh, IEEE Transactions on Magnetics **55**, 1400304 (2019).

[10]K. Kato, S. Hagi M. Hinoshita, E. Shikoh, and Y. Teki, Phys. Chem. Chem. Phys. **19**, 18845 (2017).

[11]H. Liu, J. Wang, A. Chanana, and Z. V. Vardeny, J. Appl. Phys. **125**, 142908 (2019).

[12]X. Sun, M. Gobbi, A. Bedoya-Pinto, O. Txoperena, F. Golmar, R. Llopis, A. Chuvilin, F. Casanova, and L. Hueso, Nature Commun. **4**, 2794 (2013).

[13]M. Cinchetti, K. Heimer, J.-P. Wüstenberg, O. Andreyev, M. Bauer, S. Lach, C. Ziegler, Y. Gao, and M. Aeschlimann, Nature Mater. **8**, 115 (2009).

[14]Z. Tang, S. Tanabe, D. Hatanaka, T. Nozaki, T. Shinjo, S. Mizukami, Y. Ando, Y. Suzuki, and M. Shiraishi, Jpn. J. Appl. Phys. **49** 033002 (2010).

[15]S. W. Jiang, P. Wang, B.B. Chen, Y. Zhou, H.F. Ding, and D. Wu, Appl. Phys. Lett. **107**, 042407 (2015).

[16]Y. Liu, T. Lee, H.E. Katz, and D.H. Reich, J. Appl. Phys. **105**, 07C708 (2009).

[17]J.-Y. Hong, K.-H. O. Yang, B.-Y. Wang, K-S. Li, H.-W. Shiu, C.-H. Chen, Y.-L. Chan, D.-H. Wei, F.-H. Chang, H.-J. Lin, W.-C. Chiang, and M.-T. Lin, Appl. Phys. Lett. **104**, 083301 (2014).

[18]V. Bulović, P. Tian, P. E. Burrows, M. R. Gokhale, S. R. Forrest, and M. E. Thompson, Appl. Phys. Lett. **70**, 2954 (1997).

[19]S. Mizukami, Y. Ando, and T. Miyazaki, Phys. Rev. B **66**, 104413 (2002).

[20]Y. Tserkovnyak, A. Brataas, and G.E.W. Bauer, Phys. Rev. Lett. **88**, 117601 (2001).

[21]E. Shikoh, K. Ando, K. Kubo, E. Saitoh, T. Shinjo, and M. Shiraishi, Phys. Rev. Lett. **110**, 127201 (2013).

[22]E. Saitoh, M. Ueda, H. Miyajima, and G. Tatara, Appl. Phys. Lett. **88**, 182509 (2006).

[23]K. Ando and E. Saitoh, J. Appl. Phys. **108**, 113925 (2010).